\documentclass[12pt,preprint]{aastex}

\usepackage{graphicx}
\usepackage{overpic}

\shorttitle{Fan-shaped jets}
\shortauthors{Jiang, R. L. et al.}

\begin{document}

\title{Fan-shaped jets in three dimensional reconnection simulation as a model of ubiquitous solar jets}
\author{Rong Lin Jiang\altaffilmark{1,2,4}, Kazunari Shibata\altaffilmark{2}, Hiroaki Isobe\altaffilmark{3} \& Cheng Fang\altaffilmark{1,4}}

\email{rljiang@nju.edu.cn}

\altaffiltext{1}{Department of Astronomy, Nanjing University, Nanjing 210093, China}
\altaffiltext{2}{Kwasan and Hida Observatories, Kyoto University, Yamashina, Kyoto 607-8471, Japan}
\altaffiltext{3}{Unit of synergetic Studies for Space, Kyoto University, Yamashina, Kyoto 607-8471, Japan}
\altaffiltext{4}{Key Laboratory of Modern Astronomy and Astrophysics (Nanjing University), Ministry of Education, China}

\begin{abstract}
Magnetic reconnection is a fundamental process in space and astrophysical plasmas in which oppositely directed magnetic fields changes its connectivity and eventually converts its energy into kinetic and thermal energy of the plasma. Recently, ubiquitous jets (for example, chromospheric anemone jets, penumbral microjets, umbral light bridge jets) have been observed by Solar Optical Telescope on board the satellite Hinode. These tiny and frequently occurring jets are considered to be a possible evidence of small-scale ubiquitous reconnection in the solar atmosphere. However, the details of three dimensional magnetic configuration are still not very clear. Here we propose a new model based on three dimensional simulations of magnetic reconnection using a typical current sheet magnetic configuration with a strong guide field. The most interesting feature is that the jets produced by the reconnection eventually move along the guide field lines. This model provides a fresh understanding of newly discovered ubiquitous jets and moreover a new observational basis for the theory of astrophysical magnetic reconnection.
\end{abstract}


\section{INTRODUCTION}

Recent observations have found ubiquitous plasma jets over the solar atmosphere in various plasma parameters and magnetic field configuration. The solar chromospheric anemone jets show a cusp- or inverted Y-shaped structure which are believed to be a result of magnetic reconnection between a magnetic bipole and a preexisting uniform vertical field~\citep{Shibata2007}. The penumbral microjets show the apparent motion almost along the vertical guide field components in the interlocking-comb structure of magnetic field lines in the sunspot penumbra~\citep{Katsukawa2007}. The umbral light bridge jets are ejected along the vertical field lines emanating from the light bridge in the sunspot umbra~\citep{Shimizu2009}. There are many other chromospheric jets whose footpoints are not well resolved. It has sometimes been proposed that spicules may be produced by magnetic reconnection~\citep{Suematsu2008, Isobe2008}. The three dimensional magnetic field configuration at the footpoints of these jets are still puzzling. Especially, how these jets are accelerated is a fundamental question in solar physics, which would also give a hint to the understanding of the origin of astrophysical jets. The three dimensional (3D) numerical simulation may help us to understand this. For 3D magnetic reconnection, many simulations show the reconnecting process is much more complicated and difficult than two dimensional (2D) case~\citep{Yokoyama1995, Chen1999, Chen2001, Yokoyama2001, Isobe2005, Jiang2010}. Some of the 3D simulations have shown generation of flows parallel to magnetic field lines as a result of non-null reconnection~\citep{Pontin2005, Ugai2010}. However, these parallel flows have not been analyzed in detail. In this paper, we analyzed these flows in detail for the first time, because these flows are important as the origin of chromospheric jets and found that these jets ejected from the diffusion region move along the magnetic guide field and its 3D structure is similar to a fan-shape (hereafter, referred to as fan-shaped jets) which differ from the classical reconnection theory (generally speaking, the reconnection jets move along the ambient magnetic field, hereafter, ordinary reconnection jets)~\citep{Sweet1958, Parker1957, Petschek1964, Priest2000}. In this report, we give a description and analysis for these results.

\section{NUMERICAL METHOD}

\begin{figure}
\centering
\includegraphics[width=400pt]{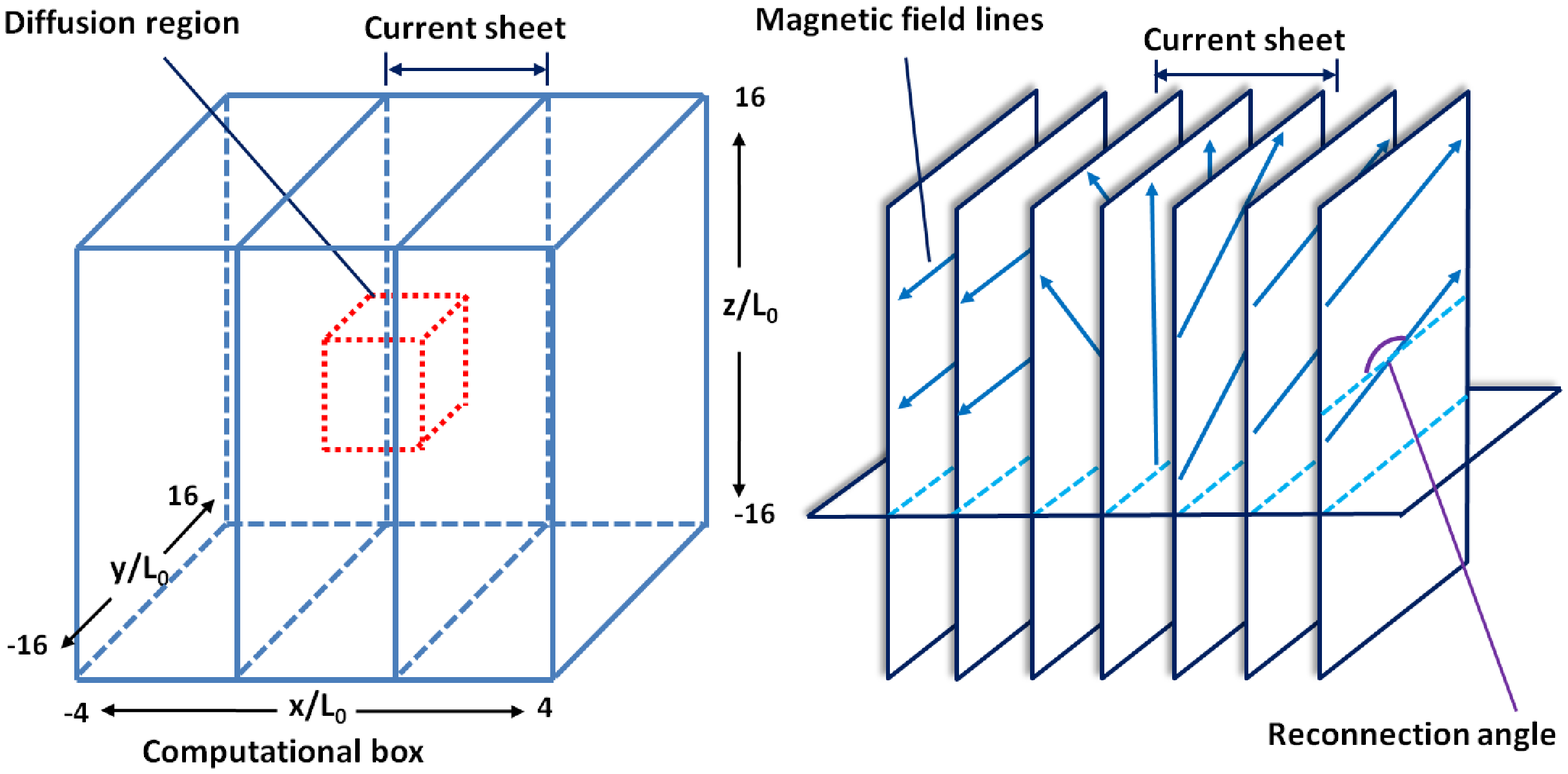}
\caption{ Initial magnetic field configuration. The box size is $-4<x<4$, $-16<y<16$, $-16<z<16$. The magnetic field lines rotate from $x = -0.4$ to $x = 0.4$. The initial uniform distributions of density and pressure are used in our simulations. The magnetic hydrostatic equilibrium is satisfied at the initial condition. The normalization units (the values used in our analysis and discussion is in nondimensional for simplicity) of length, temperature, density, pressure, time, velocity and magnetic field are $L_0$, $T_0$, $\rho_0$, $\rho_0T_0\kappa_B/m$, $L_0/v_0$, $(p_0/\rho_0)^{1/2}$ and $(p_0)^{1/2}$, respectively ($\kappa_B$ and $m$ mean Boltzmann constant and molecular mass). We assume that an anomalous resistivity, with the form $\eta = \eta_{ini} \cos ( x \pi / 2 \triangle h) \cos ( y \pi /2 \triangle h) \cos ( z \pi /2 \triangle h)$ (the $\eta_{ini}$ is the amplitude of resistivity in the diffusion region), is localized in a small region $|x| \leq \triangle h$, $|y| \leq \triangle h$, $|z| \leq \triangle h$, where $\triangle h = 0.4$. The typical magnetic Reynolds number, plasma beta and Alfv\'en speed is $126$, $0.8$ and $1.58$, respectively, where the Reynolds number is calculated using $\eta_{ini}$ (0.05), Alfv\'en speed (1.58) and the half length of the computational box in x direction (4). \label{fig1}}
\end{figure}

We perform three dimensional magnetohydrodynamic (MHD) simulation of magnetic reconnection using a typical force free magnetic field. The mathematical form of this field is

\begin{equation}
  B_x=0
\end{equation}

\begin{equation}
  B_y=\left\{
  \begin{array}{ll}
    -B_{ini}                                    & \ \ \textrm{for} \ \   x < -\triangle h_x  \\
    B_{ini}\sin{\{[\theta x - (\pi - \theta)\triangle h_x]/2\triangle h_x\}}  & \ \ \textrm{for} \ \   |x| < \triangle h_x \\
    B_{ini}\sin{[(2 \theta - \pi)/2]}           & \ \ \textrm{for} \ \   x > \triangle h_x  \\
  \end{array}
  \right.
\end{equation}

\begin{equation}
  B_z=\left\{
  \begin{array}{ll}
    0                                     & \ \ \textrm{for} \ \   x < -\triangle h_x  \\
    B_{ini}\cos{\{[\theta x - (\pi - \theta)\triangle h_x]/2\triangle h_x\}}   & \ \ \textrm{for} \ \   |x| < \triangle h_x \\
    B_{ini}\cos{[(2 \theta - \pi)/2]}     & \ \ \textrm{for} \ \   x > \triangle h_x  \\
  \end{array}
  \right.
\end{equation}

\noindent where $B_{ini}$ is the initial amplitude of magnetic field. The $\triangle h_x$ is the half width of the current sheet. $\theta$ is the reconnection angle shown in Figure~\ref{fig1}, and it can be four values in our simulation cases, i.e., $\pi$, $3\pi/4$, $\pi/2$ and $\pi/3$.

The computational box is resolved by $240 \times 480 \times 480$ grid points, as shown in Figure~\ref{fig1}. Since the magnetic field in the current sheet is a function of $x$, we use a non-uniform grid point distribution in the $x$ direction. The eight independent variables are density ($\rho$), velocity ($v_x$, $v_y$, $v_z$), magnetic field ($B_x$, $B_y$, $B_z$) and gas pressure (p). We studied four cases for different initial reconnection angles ($\pi/3$, $\pi/2$, $3\pi/4$ and $\pi$) which are defined in Figure~\ref{fig1}. Using the free boundary condition and the CIP-MOCCT numerical scheme~\citep{Yabe1991a, Yabe1991b, Kudoh1999}, we solved three dimensional compressible resistive MHD equations. Note that there is no thermal conduction and gravity being considered in our simulations because what we are really interested in is the process of magnetic reconnection.

\section{RESULTS}

\begin{figure}
\centering
\begin{overpic}[width=230pt,trim=12mm 20mm 12mm 20mm,clip]{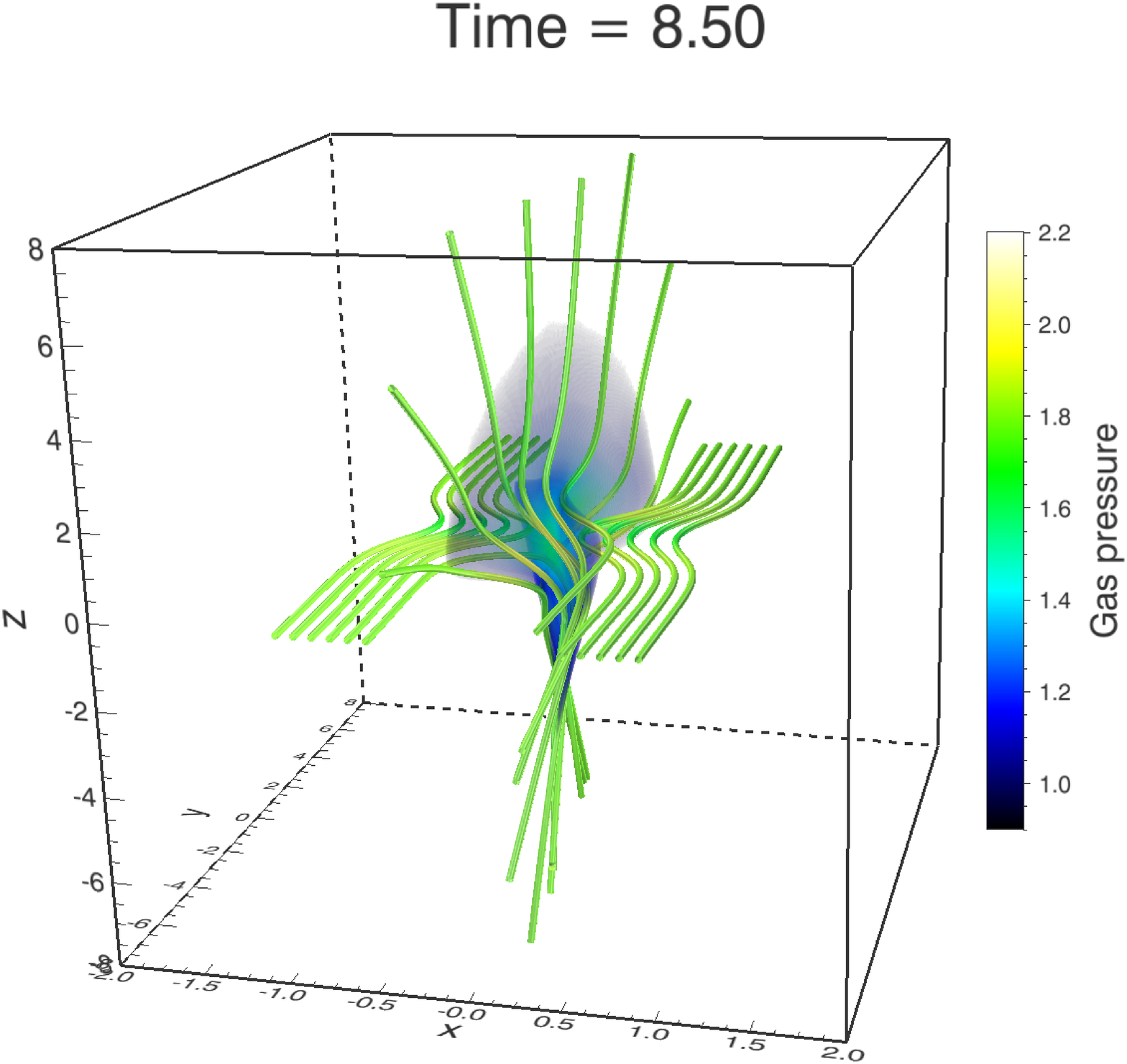}
\put(10,90){\textbf{a}}
\end{overpic}
\begin{overpic}[width=230pt,trim=12mm 20mm 12mm 20mm,clip]{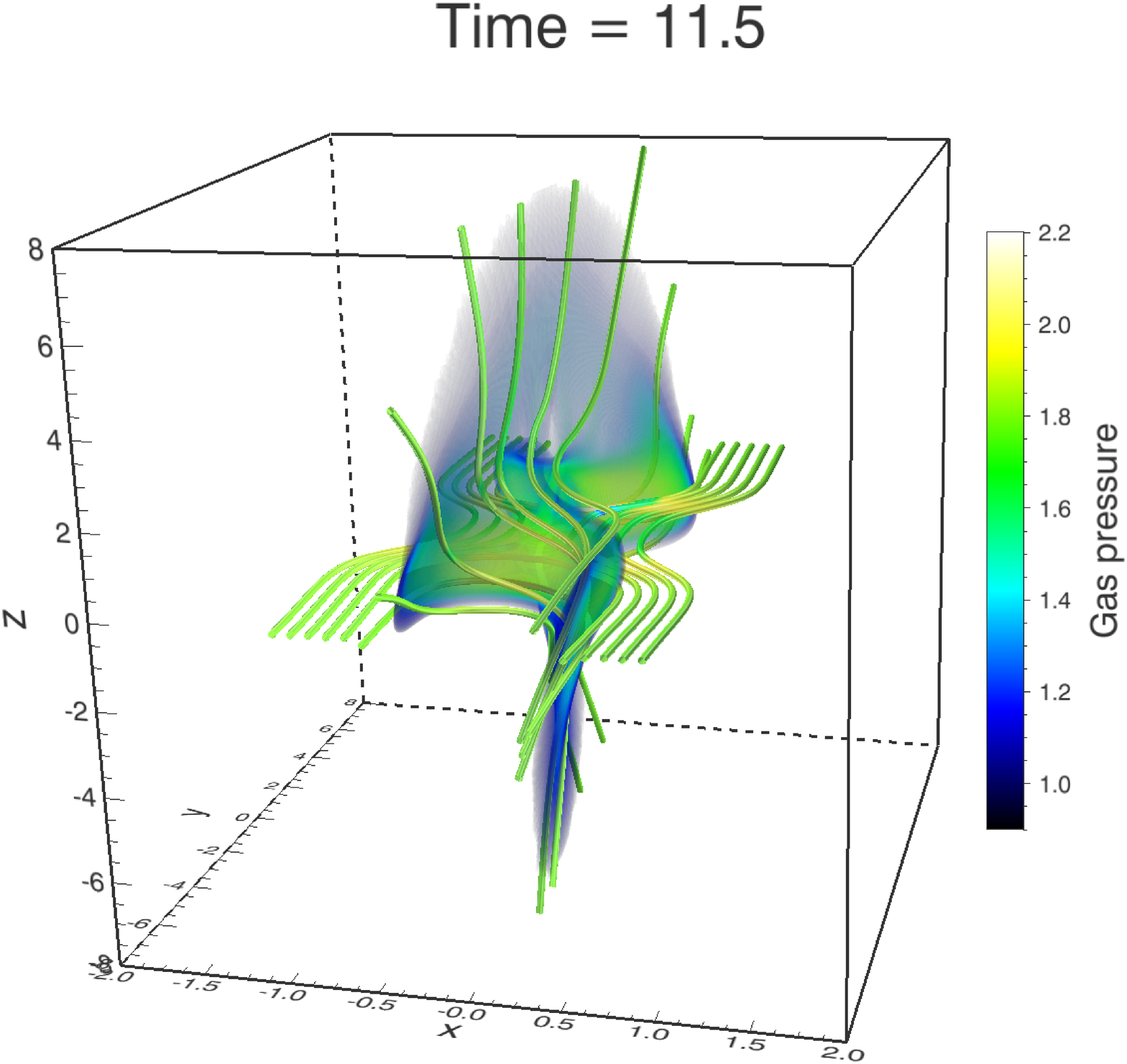}
\put(10,90){\textbf{b}}
\end{overpic}
\begin{overpic}[width=230pt,trim=12mm 20mm 12mm 20mm,clip]{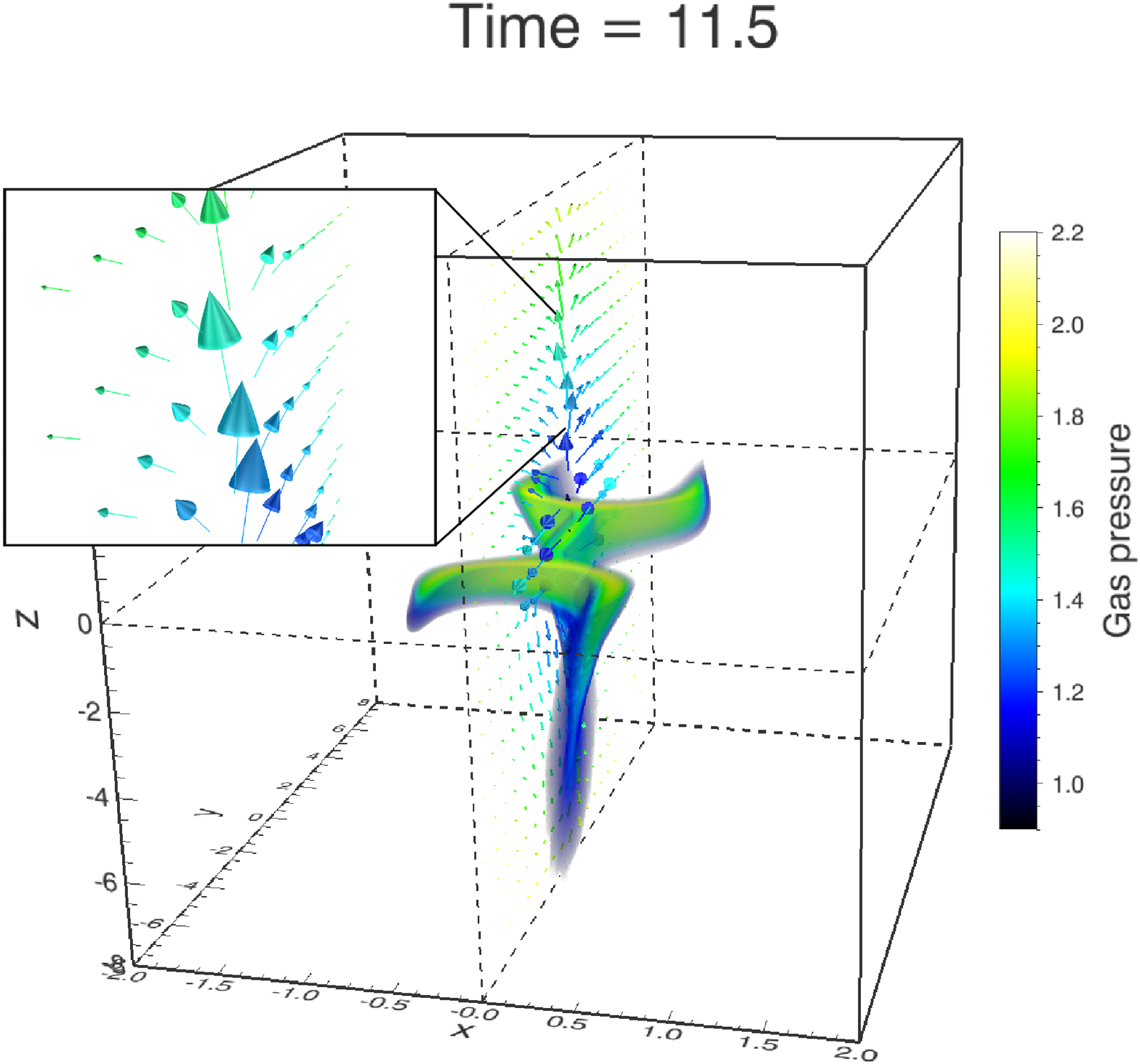}
\put(10,90){\textbf{c}}
\end{overpic}
\begin{overpic}[width=230pt,trim=12mm 20mm 12mm 20mm,clip]{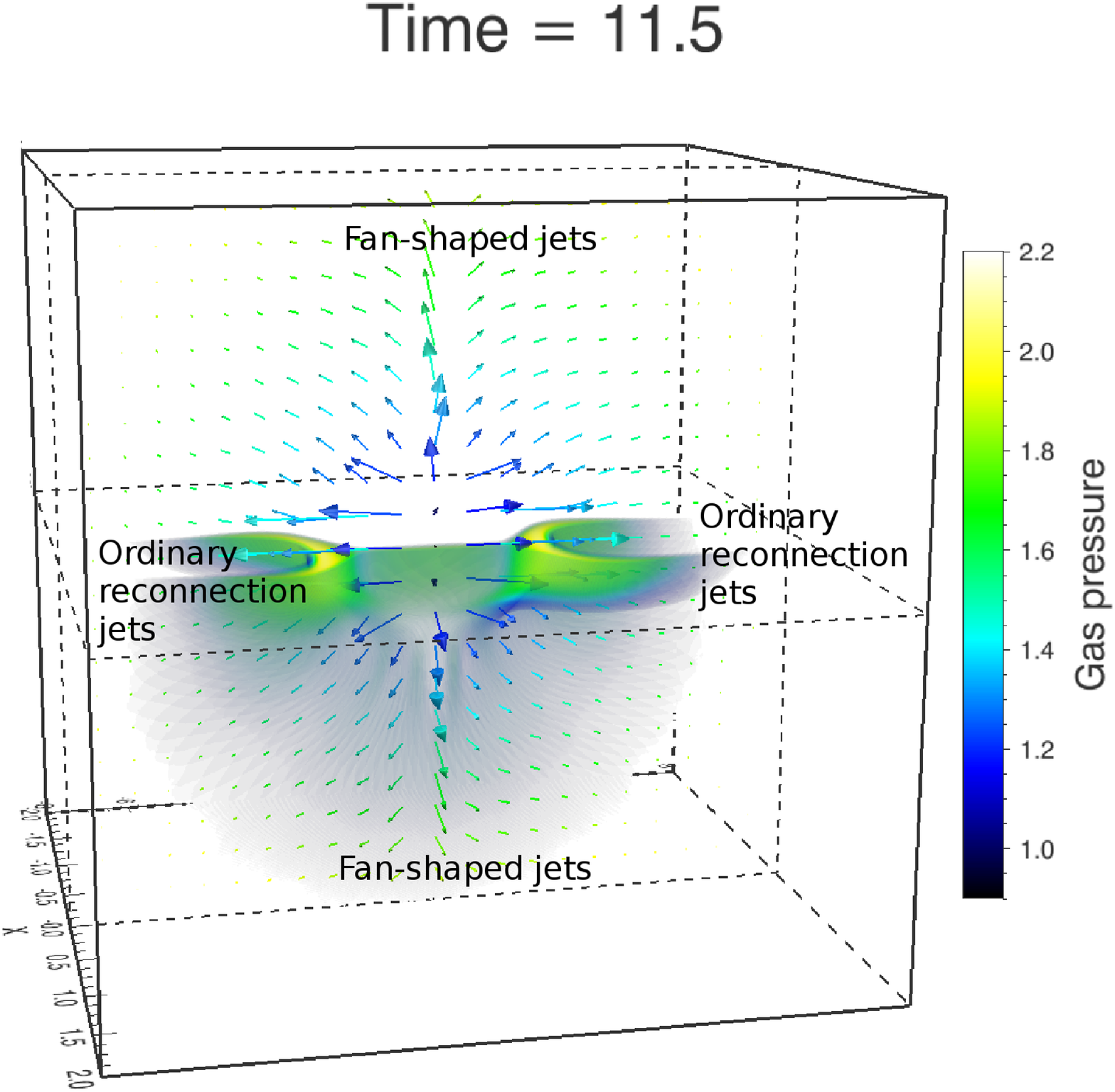}
\put(10,90){\textbf{d}}
\end{overpic}
\caption{Typical case of a fan-shaped jet. The 3D visualization (panels a-d) of the gas pressure distribution with the reconnection angle $\pi$. Solid tubes are magnetic field lines and arrows are velocity. (\textbf{a}) 3D visualization of pressure distribution at the time 8.5. (\textbf{b}) 3D visualization of pressure distribution at the time 11.5. (\textbf{c} and \textbf{d}) are the same distribution at the time 11.5, but we added a cross section at $z=0$ (the gas pressure rendering above this plane is hidden) and a velocity field plane at $x=0$. Moreover, (\textbf{d}) shows image seen from the right side of (\textbf{c}) (the view angle is different). Since the current sheet is very thin, a different scale in $x$ direction is used in these panels. \label{fig2}}
\end{figure}

Figure \ref{fig2} shows the 3D visualization of the gas pressure distribution at the $time = 8.5$ and $time = 11.5$ of our typical case (the reconnection angle is $\pi$). We added a cross section at $z=0$ and a velocity field plane at $x=0$ in the lower panels, in which we can see the X-shaped structure of ordinary reconnection outflow on the $x-y$ plane and the velocity distribution of the fan-shaped jets on the $y-z$ plane. The plasma is heated and the magnetic field lines disconnect and reconnect at central diffusion region. Then we have two kinds of jets. The one is the ordinary reconnection outflow which is shown by the X-shaped structure on the $x-y$ plane. Besides, it can be seen that there are two fan-shaped jets moving along the z and -z direction. The biggest feature of these jets is that they can move along the magnetic guide field. Since the magnetic field is shearing in the current sheet, the jets can move in different direction (the shearing magnetic field means the direction of the magnetic field lines is different if they have a different position $x$ in the diffusion region). The shape of ordinary reconnection jets on $x-y$ plane is very similar to the results in 2D reconnection case and the velocity is about the local Alfv\'en speed at the $time = 11.5$ (the speed of the fan-shape jets is about half of Alfv\'en speed).

\begin{figure}
\centering
\begin{overpic}[width=230pt]{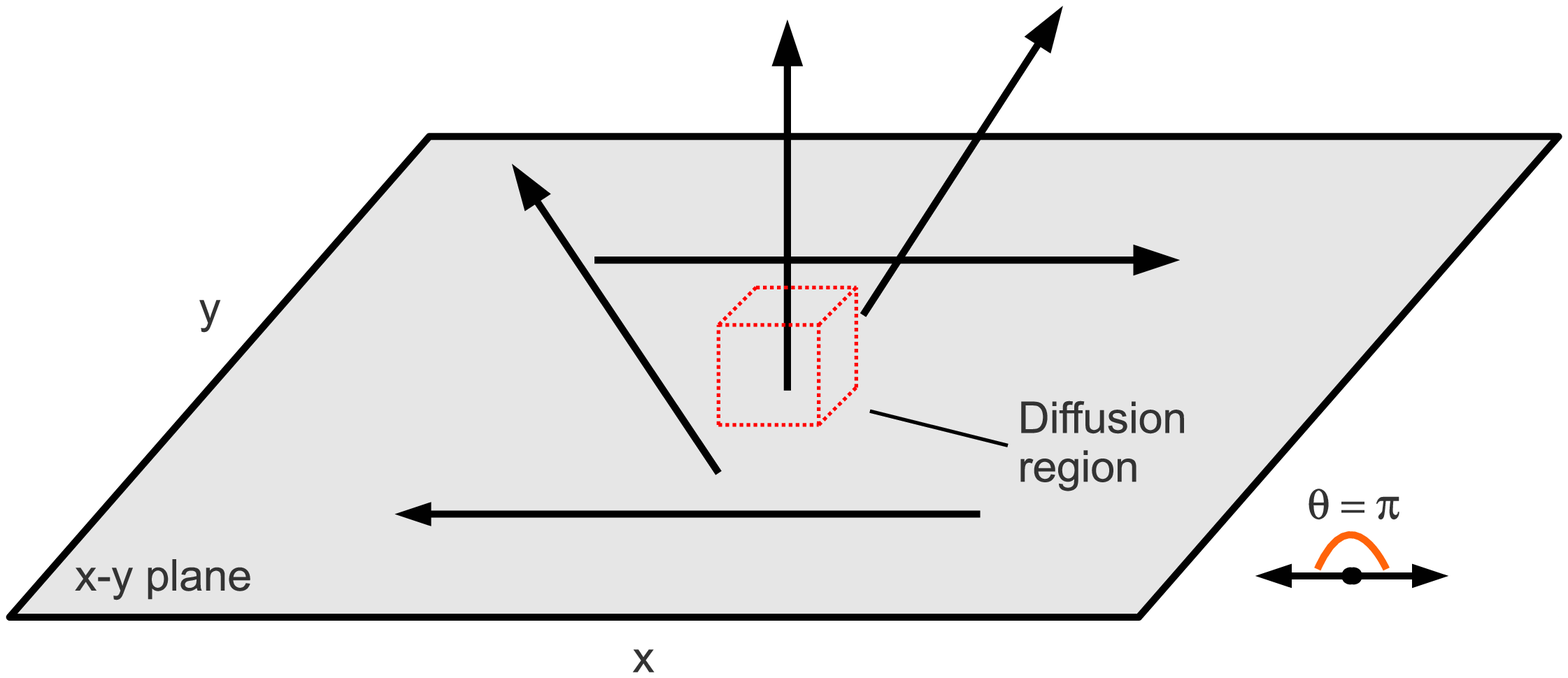}
\put(10,35){\textbf{a}}
\end{overpic}
\begin{overpic}[width=230pt]{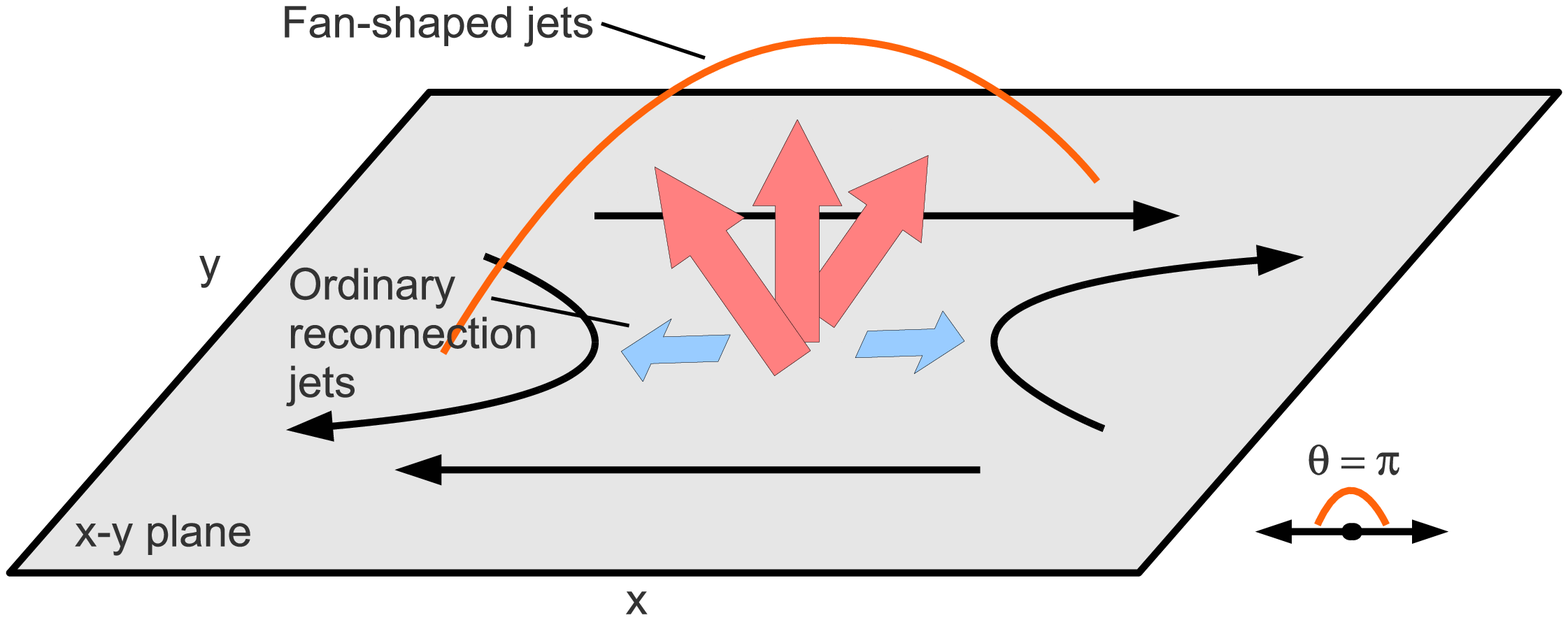}
\put(10,35){\textbf{b}}
\end{overpic}
\begin{overpic}[width=230pt]{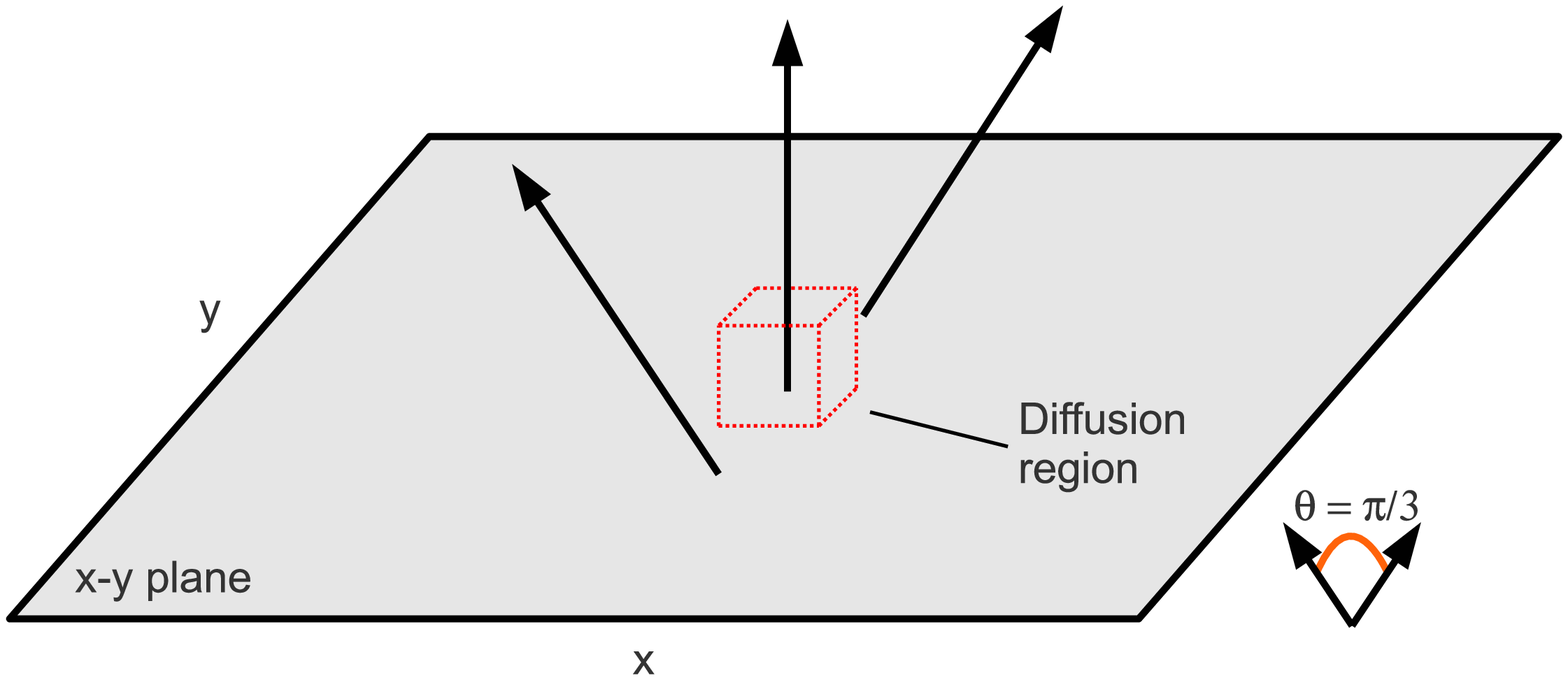}
\put(10,35){\textbf{c}}
\end{overpic}
\begin{overpic}[width=230pt]{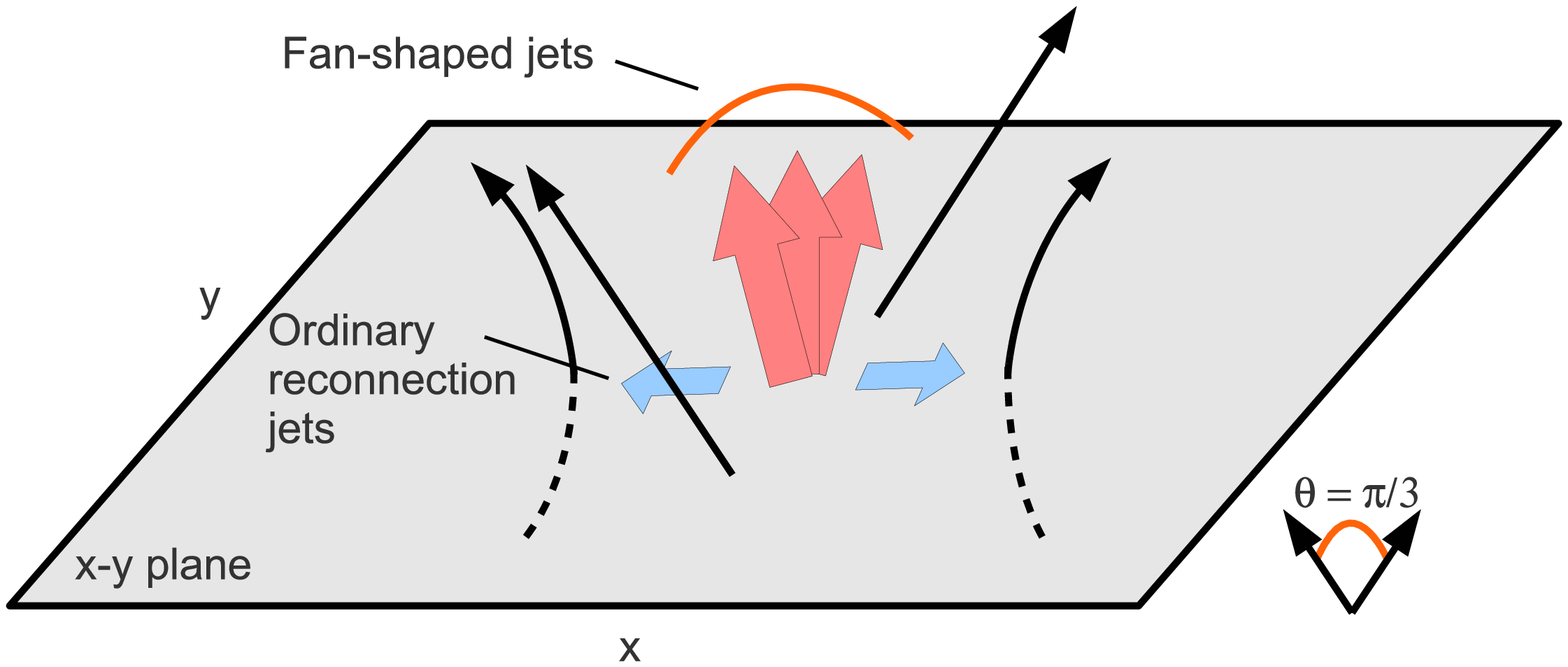}
\put(10,35){\textbf{d}}
\end{overpic}
\caption{A cartoon for explaining the fan-shaped jets, which shows both ordinary reconnection jets and the fan-shaped jets. The blue jets are the ordinary reconnection jets and the red ones are the fan-shaped jets. (\textbf{a}) The initial magnetic configuration with the reconnection angle $\pi$. (\textbf{b}) The fan-shaped and ordinary jets with the reconnection angle $\pi$. (\textbf{c} and \textbf{d}) are the same as (\textbf{a}) and (\textbf{b}) but the reconnection angle is $\pi/3$. Note that the magnetic configuration is shearing in the current sheet and the jets always move along the magnetic field lines, so the structure looks like a fan-shape. \label{fig3}}
\end{figure}

Figure~\ref{fig3} shows a cartoon for explaining the fan-shaped jets (red arrows) and the ordinary reconnection jets (blue arrows). The fan-shaped jets are ejected from the central diffusion region in different direction which is due to the shearing magnetic field in the current sheet. Finally, we have a structure which looks like a fan-shape. The reconnection angle for Figure~\ref{fig3}\textbf{a} and \textbf{b} is $\pi$, and that for Figure~\ref{fig3}\textbf{c} and \textbf{d} is $\pi/3$. As we described above, the fan-shaped jets can move along the guide field in our typical case. For the small reconnection angle case, similar to the typical case, two oblique magnetic field can produce the fan-shaped jets in a narrow angle. This situation is very general in the solar atmosphere. If a vertical magnetic flux tube is twisted or sheared or if two magnetic field lines or tubes with different directions get close to each other (as shown in the Figure~\ref{fig3}\textbf{c} and \textbf{d} considering $x-y$ plane as the solar surface), large electric currents are generated between them, resulting in fast magnetic reconnection with larger amount of energy released. The ordinary reconnection jets and fan-shaped jets are ejected from the diffusion region simultaneously. Since the ordinary reconnection jets almost parallel to the solar surface, they can only move a short distance and disappear. However, the fan-shaped jets can move upwards and become the jets along the vertical field lines. It should be noted that the density stratification effect in the solar atmosphere can greatly increase velocity amplitude of the slow mode shock ahead of the upward fan-shaped jets when it propagates to the low density region (like the upper chromosphere and solar corona) and eventually become a faster jet accelerated by the slow shock~\citep{Shibata1982a}.

\begin{figure}
\centering
\begin{overpic}[width=170pt,trim=25mm 15mm 25mm 25mm,clip]{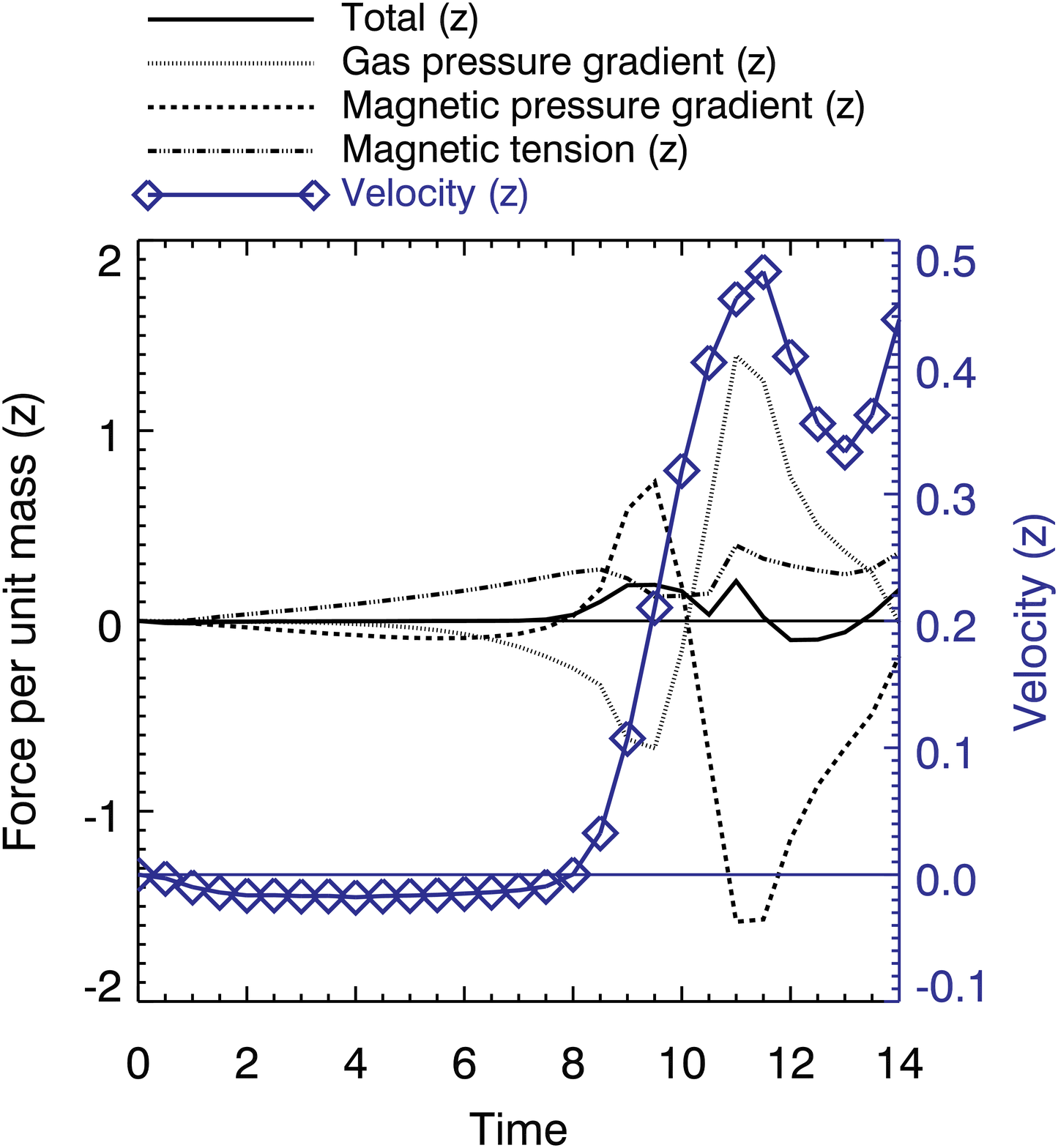}
\put(2,90){\textbf{a}}
\end{overpic}
\begin{overpic}[width=140pt,trim=20mm 20mm 25mm 40mm,clip]{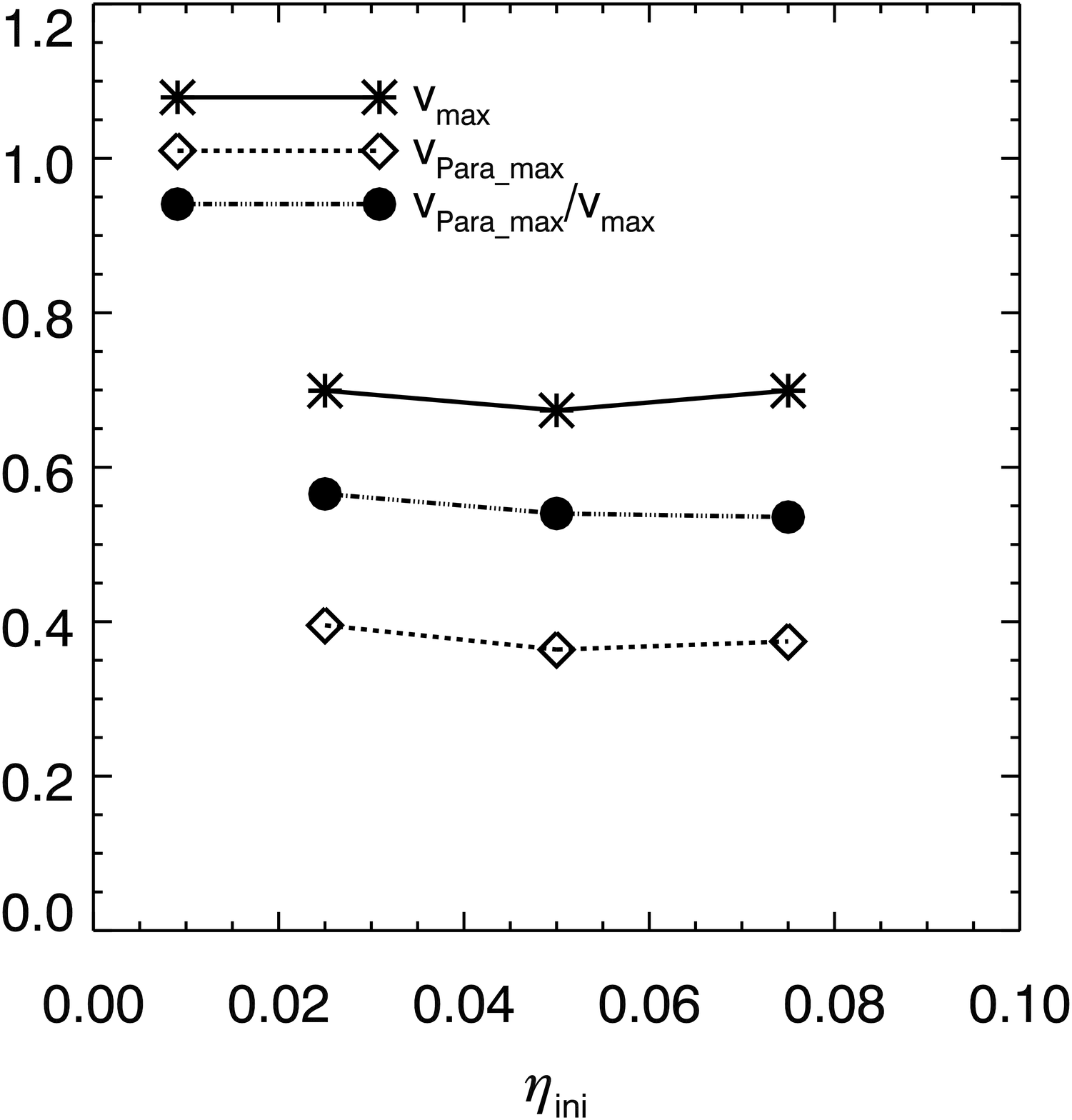}
\put(2,95){\textbf{b}}
\end{overpic}
\begin{overpic}[width=140pt,trim=20mm 20mm 25mm 40mm,clip]{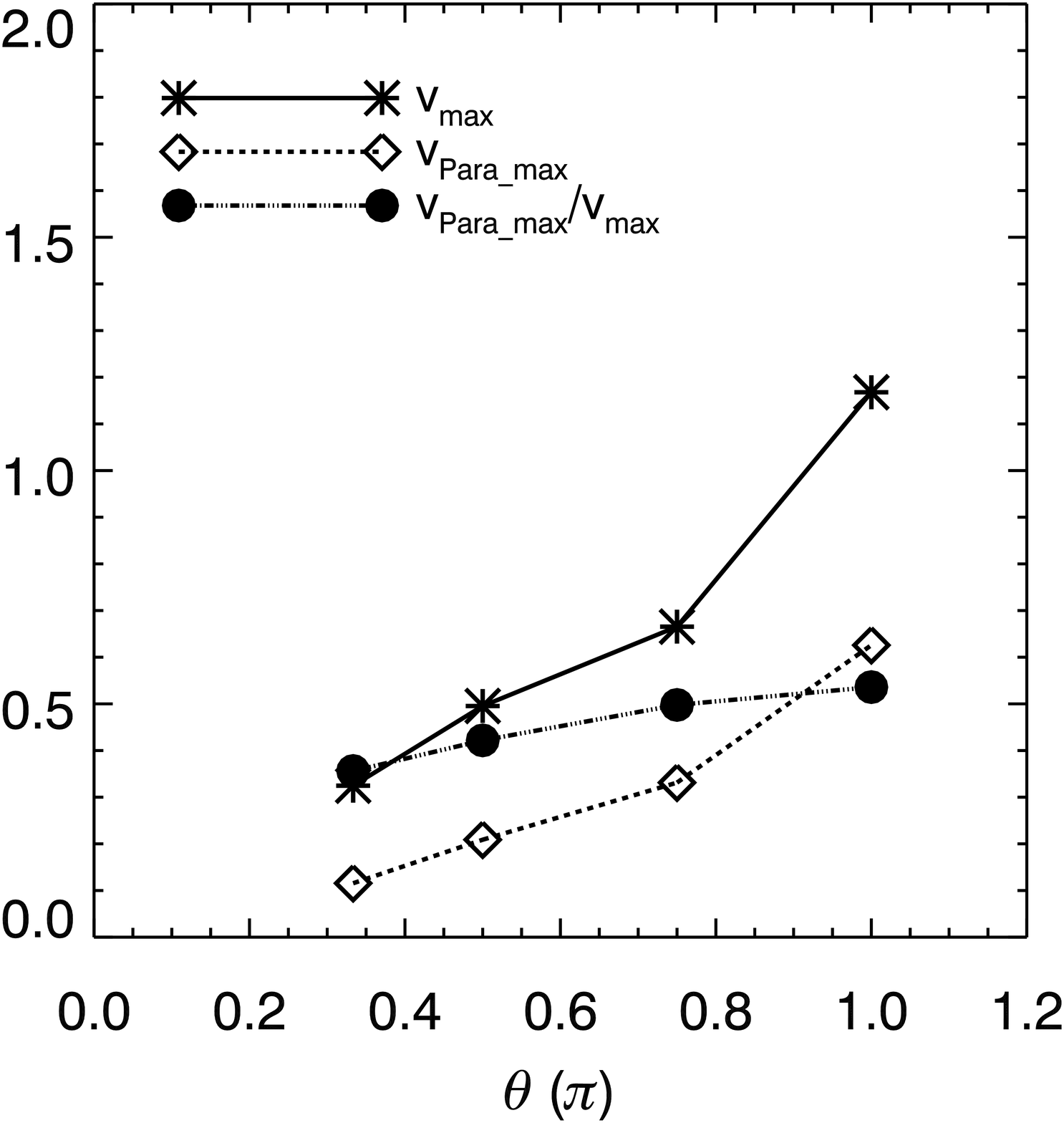}
\put(2,95){\textbf{c}}
\end{overpic}
\caption{(\textbf{a}) This panel shows the forces per unit mass and the velocity of one Lagrangian fluid element in the z-direction for this typical case. The other panels show the dependence of the jet velocity on the resistivity and the reconnection angle. The value $v_{max}$ is the maximum velocity in the computational box which indicates the speed of ordinary reconnection jets. $v_{Para\_max}$ is the maximum velocity along the initial magnetic field line which means the velocity of fan-shaped jets. (\textbf{b}) This is the dependence of the jet velocity on the resistivity where the $\eta_{ini}$ is the amplitude of resistivity in the diffusion region. For this case we choose three resistivity value 0.025, 0.05 and 0.075 with the plasma beta 1.0. (\textbf{c}) This panel shows the dependence of the jet velocity on the reconnection angle. For this case we choose four reconnection angle value $\pi/3$, $\pi/2$, $3\pi/4$ and $\pi$ with the plasma beta 0.8. \label{fig4}}
\end{figure}

In order to understand why the jets move along the guide field, we investigate the driving force of the fan-shaped jet using Lagrangian fluid elements (test particles). These elements can move with the plasma but has no effect to the result of simulations, which can show us the detailed information at the position of these elements in the computational box. One of these fluid elements is located at the edge of the diffusion region at the initial time and finally the element is driven as a part of the fan-shaped outflow. Its forces and velocity is shown by Figure~\ref{fig4}\textbf{a}. There are two stages for acceleration. In the first stage ($time=8-10$), the Lorentz force (including magnetic tension force and magnetic pressure gradient force) drives this element. After that it is dominated by the gas pressure gradient in the later stage ($time=10-11.5$). Hence, we conclude that the fan-shaped jets are accelerated by both gas pressure gradient force and the Lorentz force. That is different from the ordinary reconnection outflow which is only accelerated by the Lorentz force.

Figure~\ref{fig4}\textbf{b} and \textbf{c} shows the dependence of velocity on the resistivity and the reconnection angle. All the velocity is measured at the time when reconnection rate reaches to the maximum for different cases. As shown in Figures~\ref{fig4}\textbf{b}, velocity are almost constant with different resistivity value. That is to say, there is no remarkable dependence of the velocity on the resistivity value. The ratio of the velocity between two kinds of jets is approximate 0.5 (the velocity of the ordinary reconnection jets is comparable to the local Alfv\'en speed determined by the reconnecting component), which indicates that the mechanism for forming these two jets is different as we described above. In Figure~\ref{fig4}\textbf{c}, the velocity is sensitive to the reconnection angle. The large reconnection angle means a strong shearing magnetic field lines in the current sheet. If we use the reconnection angle $\theta=0$ as an extreme case, there will be no reconnection process and no jets come out. Furthermore, we found that the maximum speed of fan-shaped jets is still around half of the ordinary ones.

\section{DISCUSSION}

Let us briefly discuss the application of our results to various jets in the chromosphere as shown in Table~\ref{tab1}. Although the footpoints of these jets are not necessarily well resolved, we assume that the 3D reconnection occurs in the photosphere or in the low chromosphere at the footpoints of these jets, and fan-shaped jets are ejected from the reconnection region along magnetic field lines in these layers.  The Alfv\'en speed ($v_A$) in the photosphere and low chromosphere is about 10 km s$^{-1}$ in a typical isolated flux tube outside sunspots, and is about 10-50 km s$^{-1}$ in sunspot umbra and penumbra. Hence the $v_A$ in Table~\ref{tab1} shows such local Alfv\'en speed in the (hypothetical) reconnection region at the footpoint of these jets. Here, the $v_{A\bot}$ is the Alfv\'en speed based on the reconnecting component of magnetic field (we assume that $v_{A\bot} = 0.2 v_A$). The velocity of fan-shaped jets is only half of that of the ordinary reconnection flow ($v_{fan} \sim v_{A\bot}/2$). Once the fan-shaped jets are ejected, the slow mode MHD shock is formed ahead of the jets and propagate along the vertical magnetic field lines. Since the density decreases with height, the velocity amplitude at the slow mode shock increases with height, namely $v_{fan,max} \sim v_{fan}e^{0.5z/H}$ if the slow mode wave energy is conserved or $v_{fan,max} \sim v_{fan}e^{0.23z/H}$ if the shock is strong~\citep{Shibata1982b}, where the value z is the height of the jets measured from the reconnection region and H is the pressure scale height. If z/H = 13.3 (assuming the scale height H is 150 km and the slow mode shock propagates over a height of 2000 km), we get $v_{fan,max} \sim v_{fan}e^{0.23z/H} \sim v_{fan}e^{3.0} \sim 20v_{fan}$. Similar results are obtained also for the case z = 1000 km when the wave energy is conserved ($v_{fan,max} \sim v_{fan}e^{0.5z/H}$). Table~\ref{tab1} shows that the resulting velocity at the shock front ($v_{fan,max}$) when it reaches at the top of the chromosphere is comparable to the actual observed velocity of these chromospheric jets. Hence, our finding of fan shaped jets parallel to field lines are important for understanding the origin of chromospheric jets and seems to be successfully applicable to ubiquitous chromospheric jets.

\begin{table}[ht]
\caption{Comparison between observation and fan-shaped jets (unit: $km~s^{-1}$).\label{tab1}}
\begin{tabular}{cccccc}
\hline
\hline
Jets & Observational velocity & $v_A$ & $v_{A \bot} $ & $v_{fan}$ & $v_{fan,max}$ \\
\hline
Chromospheric anemone jets\tablenotemark{1}    & $\sim10$    & $10$        & $2$      &  $1$     & $20$         \\
Spicules\tablenotemark{2}                      & $\sim25$    & $10$        & $2$      &  $1$     & $20$         \\
Penumbral jets\tablenotemark{3}                & $50-150$    & $10-50$     & $2-10$   &  $1-5$   & $20-100$     \\
Umbral light bright jets\tablenotemark{4}      & $28-180$    & $10-50$     & $2-10$   &  $1-5$   & $20-100$     \\
\hline
\end{tabular}
\tablenotetext{1}{\cite{Shibata2007}}
\tablenotetext{2}{\cite{Suematsu2008}}
\tablenotetext{3}{\cite{Katsukawa2007}}
\tablenotetext{4}{\cite{Shimizu2009}}
\end{table}

We described fan-shaped jets by simulating the 3D reconnection process using a simple initial shearing magnetic configuration in this report. We found that the fan-shaped jets which are accelerated by both gas pressure gradient and Lorentz force can move along the magnetic guide field lines and the velocity of these jets is about half of the local Alfv\'en speed determined by the reconnecting component of magnetic field. This new finding provides us a new way to understanding the magnetic reconnection in 3D geometry and it is also a new model for explaining the solar ubiquitous chromospheric or more general astrophysical jets. The details of them will be studied in our future papers.

\acknowledgments We thank E. Asano, A. Hillier and K. Nishida for helpful discussions. This work was supported by the National Natural Science Foundation of China (NSFC) under the grant numbers 10878002, 10610099, 10933003 and 10673004, as well as the grant from the 973 project 2011CB811402 of China, and in part by the Grant-in-Aid for Creative Scientific Research ``The Basic Study of Space Weather Prediction'' (Head Investigator: K. Shibata) from the Ministry of Education, Culture, Sports, Science and Technology (MEXT) of Japan, and in part by the Grand-in-Aid for the Global COE program ``The Next Generation of Physics, Spun from Universality and Emergence'' from MEXT. Numerical computations were carried out on Cray XT4 at Center for Computational Astrophysics, CfCA, of National Astronomical Observatory of Japan, and in part performed with the support and under the auspices of the NIFS Collaboration Research program (NIFS07KTBL005), and in part performed with the KDK system of Research Institute for Sustainable Humanosphere (RISH) at Kyoto University as a collaborative research project.

\clearpage

\clearpage

\clearpage

\end{document}